\documentstyle[12pt,a4]{article}
\input epsf

\global\arraycolsep=2pt
\def\case#1#2{{\textstyle\frac{#1}{#2}}}

\begin{document}

\begin{titlepage}

\begin{flushright}
CERN-TH/96-238\\
Budker INP 97-1\\
hep-ph/9701262
\end{flushright}

\vspace{0.5cm}
\begin{center}
\Large\bf Hybrid Renormalization of Penguins\\
and Dimension-5 Heavy--Light Operators
\end{center}

\vspace{1.5cm}
\begin{center}
Andrey G. Grozin\\
{\sl Budker Institute of Nuclear Physics, Novosibirsk 630090,
Russia\\}
\vspace{0.5cm}
and\\
\vspace{0.5cm}
Matthias Neubert\\
{\sl Theory Division, CERN, CH-1211 Geneva 23, Switzerland}
\end{center}

\vspace{1.0cm}
\begin{abstract}
We discuss the renormalization of local, scalar and pseudoscalar
dimension-5 operators containing a heavy and a light quark field at
scales below the heavy-quark mass, using the formalism of the
heavy-quark effective theory (HQET). We calculate the anomalous
dimensions of these operators and their mixing to one-loop order. We
also perform the one-loop matching of gluon and photon penguin
operators onto operators of the HQET. We discuss applications of our
results to the mixing of gluon and photon penguin operators at low
renormalization scales, and to the calculation of $1/m_Q^2$
corrections to meson decays constants.
\end{abstract}

\vspace{1.0cm}
\centerline{(Submitted to Nuclear Physics B)}

\vspace{2.5cm}
\noindent
CERN-TH/96-238\\
December 1996

\end{titlepage}

\section{Introduction}

Local operators containing both heavy and light quark fields exhibit
an interesting behaviour under renormalization at scales below the
heavy-quark mass $m_Q$. Then there arise large logarithms of the
type $\alpha_s\ln(m_Q/\mu)$, characterizing the exchange of gluons
that are ``hard'' with respect to the light quark but ``soft'' with
respect to the heavy quark. To leading order in an expansion in
$1/m_Q$, such gluons see the heavy quark as a static colour source.
The large logarithms can thus be summed to all orders in perturbation
theory using an effective theory for static heavy quarks, the
so-called heavy-quark effective theory (HQET)
\cite{EiHi}--\cite{review}. In the HQET, the 4-component heavy-quark
field $Q(x)$ is replaced by a velocity-dependent 2-component field
$h_v(x)$ satisfying $\rlap{/}v\,h_v=h_v$, where $v$ is the velocity
of the hadron containing the heavy quark. Because of the particular
hierarchy of mass scales involved, the renormalization of
heavy--light operators in the HQET is called ``hybrid''
renormalization. Operators in the effective theory have a different
evolution than in usual QCD. For instance, whereas the vector current
$\bar q\,\gamma^\mu Q$ is conserved in QCD (i.e.\ its anomalous
dimension vanishes), the corresponding current $\bar q\,\gamma^\mu
h_v$ in the HQET has a non-trivial anomalous dimension
\cite{KoRa}--\cite{PoWi}, which governs the evolution for scales
below the heavy-quark mass.

In the literature, hybrid renormalization has been discussed
extensively for local current operators of dimension 3
\cite{KoRa}--\cite{Gim} and 4 \cite{FaGr,MN94}, as well as for
4-quark operators such as the ones governing $B$--$\bar B$ mixing
\cite{ShiV}--\cite{Chris}. Here we shall consider the renormalization
of local dimension-5 operators in the HQET. The matrix elements of
such operators appear, for instance, at order $1/m_Q^2$ in the
heavy-quark expansion of weak transition form factors. In particular,
they contribute to the decay constants of heavy mesons, which have
been explored already in great detail at leading and next-to-leading
order in $1/m_Q$ \cite{subl}. The same operator matrix elements also
determine certain moments of meson wave functions \cite{GrNe}, which
play an important role in the heavy-flavour phenomenology. The
theoretical predictions for weak decay form factors involve operator
matrix elements renormalized at the scale $m_Q$. Our results can then
be used to rewrite these matrix elements in terms of ones
renormalized at a scale $\mu\ll m_Q$, which may be identified with
the scale at which a non-perturbative evaluation of these matrix
elements is performed (such as the inverse lattice spacing in lattice
field theory, or the Borel parameter in QCD sum rules).

Our results also apply to the hybrid renormalization of genuine
dimension-5 operators in QCD, such as the gluon and photon penguin
operators, which appear in the weak effective Hamiltonian
renormalized at the scale $m_b$ \cite{BBL}:\footnote{Our definition
of the coefficients $c_g$ and $c_\gamma$ contains a power of the
$b$-quark mass, which is usually included in the definition of the
operators.}
\begin{equation}
   {\cal H}_{\rm eff} = c_g(m_b)\,g_s\,\bar s\,(1+\gamma_5)
   \sigma_{\mu\nu} G^{\mu\nu} b + c_\gamma(m_b)\,
   e\,\bar s\,(1+\gamma_5)\sigma_{\mu\nu} F^{\mu\nu} b + \dots \,.
\label{Heff}
\end{equation}
We will derive the effective Hamiltonian at a lower scale $\mu<m_b$,
which is relevant to the calculation of hadronic matrix elements of
the penguin operators. Lattice calculations of such matrix elements,
in particular, are usually performed in the static theory (HQET),
since the $b$ quark is too heavy to be described as a dynamical field
on present-day lattices. It must be stressed, however, that the
validity of the effective theory is restricted to the kinematic
region where, in the rest frame of the heavy hadron, the light quarks
and gluons carry momenta much smaller than $m_Q$. For two-body decays
such as $B\to K^*\gamma$, one thus needs to evaluate the hadronic
matrix elements for unphysical particle momenta (i.e., $|{\bf
p}_{K^*}|<O(1~\mbox{GeV})$ rather than the physical value $|{\bf
p}_{K^*}|=\frac 12(m_B^2-m_{K^*}^2)/m_B\simeq 2.56$~GeV) and then
continue the results to the physical region. In more complicated
processes such as $B\to X_s\gamma^*\to X_s\,\ell^+\ell^-$ this
restriction no longer applies, and our results are of direct
relevance to the region where the strange particle carries a small
momentum.

In Sec.~\ref{sec:2}, we calculate the one-loop anomalous dimension
matrix of local dimension-5 heavy--light operators carrying zero
total momentum. Using the equations of motion, the problem is reduced
to the mixing of two operators containing the gluon field-strength
tensor. We solve the renormalization-group equation (RGE) for the
scale dependence of these operators in the leading logarithmic
approximation. In Sec.~\ref{sec:3}, we extend the basis to the
general case where the operators carry non-zero momentum. Then there
appear four additional operators, which can be written as the total
derivatives of some lower-dimensional operators. We show that, with a
suitable choice of the basis, there is no mixing between these new
operators and the ones considered in Sec.~\ref{sec:2}. In
Sec.~\ref{sec:4}, we extend the basis further by including operators
containing the photon field, and we calculate the mixing between
gluon and photon operators under hybrid renormalization. This extends
the analysis of the mixing of gluon and photon penguin operators,
which has been discussed previously for scales larger than the
$b$-quark mass \cite{GSW}--\cite{Misi}, to low renormalization
scales. In Sec.~\ref{sec:5}, we calculate the one-loop matching of
the QCD penguin operators onto their HQET counterparts. This provides
a test of our results for the hybrid anomalous dimensions. In
addition, the exact one-loop expressions for the Wilson coefficient
functions may be more appropriate to use than the leading-order
renormalization-group improved results in cases where $\ln(m_Q/\mu)$
is not a particularly large parameter; at least, they provide an
estimate of the importance of next-to-leading corrections. Also, the
one-loop matching conditions at the scale $m_Q$ will eventually be
part of a full next-to-leading order calculation, once the two-loop
anomalous dimensions of the operators will have been calculated. In
Sec.~\ref{sec:6}, we apply our results to the analysis of
higher-order corrections to meson decay constants, and to the
discussion of moments of meson wave functions. Section~\ref{sec:7}
contains the conclusions.

\section{Anomalous dimensions}
\label{sec:2}

We start by considering local, Lorentz-scalar operators of dimension
5, carrying zero total momentum. A basis of such operators in the
HQET can be constructed by considering operators containing two
covariant derivatives acting on the heavy-quark field. It is
convenient to adopt the background-field formalism \cite{Abot}, so
that it suffices to consider gauge-invariant operators. They are of
the form $\bar q\,\Gamma_{\mu\nu} iD^\mu iD^\nu h_v$, where
$\Gamma_{\mu\nu}\in \{v_\mu v_\nu, g_{\mu\nu}, \gamma_\mu v_\nu,
\gamma_\nu v_\mu, [\gamma_\mu,\gamma_\nu]\}$. We define:
\begin{eqnarray}
   O_1 &=& g_s\,\bar q\,\sigma_{\mu\nu} G^{\mu\nu} h_v \,,
    \nonumber\\
   O_2 &=& g_s\,\bar q\,\gamma_\mu v_\nu\,iG^{\mu\nu} h_v \,,
    \nonumber\\
   O_3 &=& \bar q\,(iv\!\cdot\!D)^2 h_v \,, \nonumber\\
   O_4 &=& \bar q\,(iD)^2 h_v \,, \nonumber\\
   O_5 &=& \bar q\,[i\rlap{\,/}{D}\,iv\!\cdot\!D
    + iv\!\cdot\!D\,i\rlap{\,/}{D}]\,h_v \,,
\label{ops}
\end{eqnarray}
where $ig_s G^{\mu\nu}=[iD^\mu,iD^\nu]$ is the gluon field-strength
tensor, and $\sigma_{\mu\nu}=\frac i2[\gamma_\mu,\gamma_\nu]$. A
basis of pseudoscalar operators $O_i^{(5)}$ can be obtained by
replacing $\bar q\to\bar q\,\gamma_5$. As long as we neglect the mass
of the light quark and work in a regularization scheme with
anticommuting $\gamma_5$, nothing changes by this replacement. Below,
we shall thus consider scalar operators, but it is understood that
the light-quark field $q$ could be replaced by $\gamma_5 q$ or,
equivalently, by a left- or right-handed field, $q_L$ or $q_R$.

Not all of the operators in (\ref{ops}) are independent when the
equations of motion, $\bar q\,(i\rlap{\,/}{D})^\dagger=0$ and
$iv\!\cdot\!D\,h_v=0$, are used.\footnote{We use the notation
$iD=i\partial+g_s A$ and $(iD)^\dagger=-i\overleftarrow{\partial}
+g_s A$.}
They give the relations $O_3\,\,\widehat =\,\,0$, $O_5\,\,\widehat
=\,-O_2$, and
\begin{eqnarray}
   O_4 &=& \bar q \left( \big[ (i\rlap{\,/}{D})^\dagger \big]^2
    - \case 12 g_s\sigma_{\mu\nu} G^{\mu\nu} \right) h_v
    - (i\partial)^2 (\bar q\,h_v) + 2 i\partial_\mu
    (\bar q\,iD^\mu h_v) \nonumber\\
   &\,\widehat =& -\frac 12\,O_1 + \hbox{total derivatives,}
\label{eom}
\end{eqnarray}
implying that for zero total momentum all operators can be reduced to
the operators $O_1$ and $O_2$. That it is legitimate to use the
equations of motion to reduce the operator basis has been established
in Ref.~\cite{Simm}.

\begin{figure}[htb]
\epsfxsize=12cm
\centerline{\epsffile{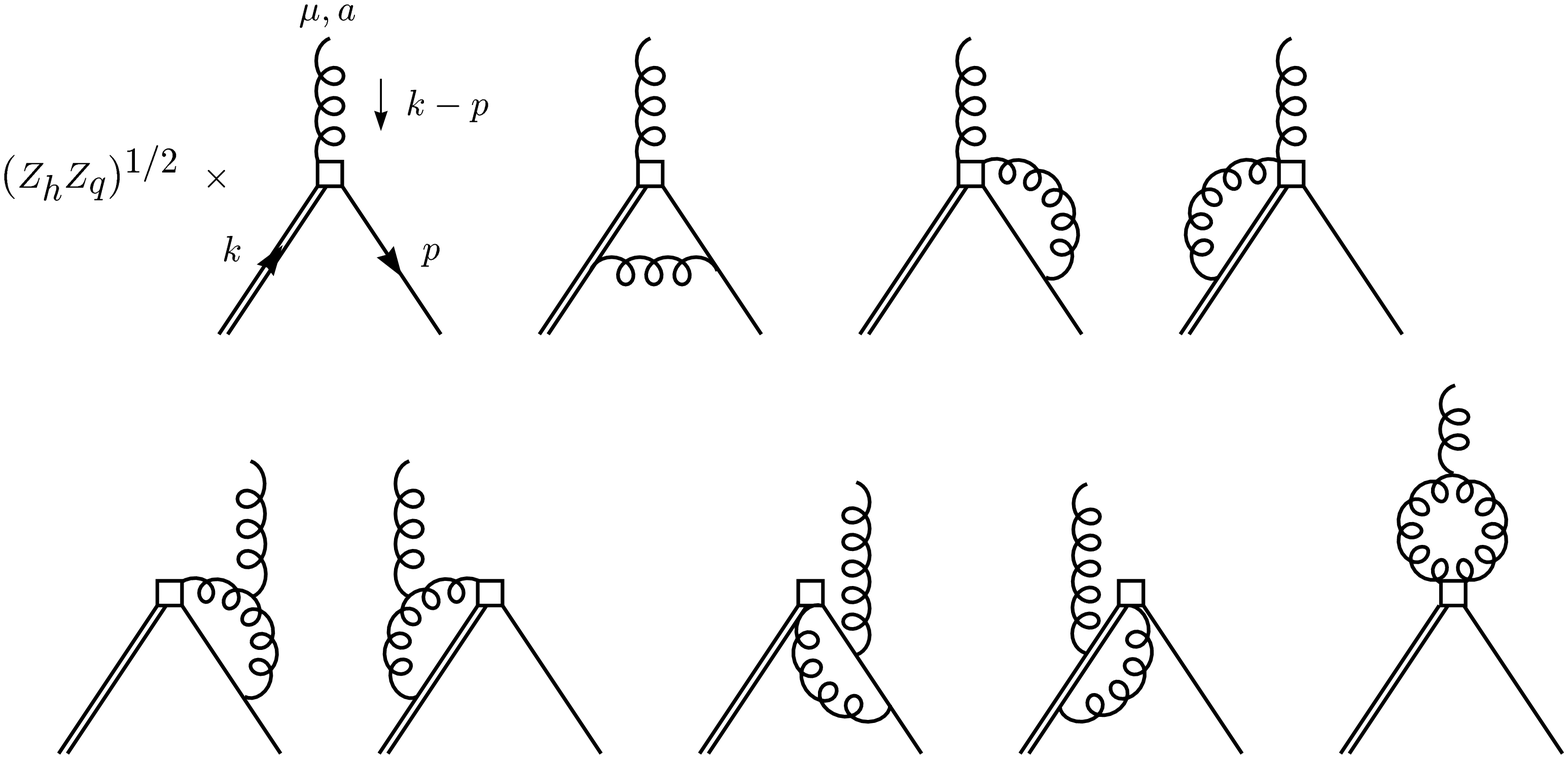}}
\centerline{\parbox{14cm}{\caption{\label{fig:diagrams}
One-loop diagrams contributing to the calculation of the anomalous
dimensions of the operators $O_1$ and $O_2$, denoted by the square.
Heavy-quark propagators in the HQET are drawn as double lines.}}}
\end{figure}

The scale dependence of the renormalized operators $O_1$ and $O_2$ is
governed by a $2\times 2$ anomalous dimension matrix, which can be
obtained by calculating the $1/\varepsilon$ poles of the matrix
elements of the bare operators in dimensional regularization (i.e.\
in $d=4-2\varepsilon$ space-time dimensions). The relevant one-loop
diagrams are shown in Fig.~\ref{fig:diagrams}. $Z_h$ and $Z_q$ are
the wave-function renormalization constants for the quark fields. In
a covariant gauge ($a=1$ corresponds to the Feynman gauge, $a=0$
corresponds to the Landau gauge), they are given by
\begin{equation}
   Z_h = 1 + C_F\,(3-a)\,\frac{\alpha_s}{4\pi\varepsilon} \,, \qquad
   Z_q = 1 - C_F\,a\,\frac{\alpha_s}{4\pi\varepsilon} \,.
\end{equation}
A virtue of the background-field formalism is that the gluon field is
not renormalized, since $Z_g Z_A^{1/2}=1$ \cite{Abot}. We have
calculated the diagrams in Fig.~\ref{fig:diagrams} in an arbitrary
covariant gauge, and with arbitrary momentum assignments (however,
for zero total momentum). The sum of all diagrams is gauge
independent, and the dependence on the external momenta combines in
such a way that the result can be expressed in terms of the matrix
elements of the basis operators in (\ref{ops}). We find
\begin{eqnarray}
   \langle O_1^{\rm bare}\rangle &=& \left\{ 1
    + \frac{\alpha_s}{4\pi\varepsilon} \left( C_F - \frac{N}{2}
    \right) \right\} \langle O_1\rangle
    + \frac{\alpha_s}{4\pi\varepsilon} \left(
    N\,\langle O_2\rangle
    + 3 C_F\,\langle O_4\rangle \right) \,, \nonumber\\
   \langle O_2^{\rm bare}\rangle &=& \left\{ 1
    + \frac{\alpha_s}{4\pi\varepsilon} \left( C_F - \frac{3N}{2}
    \right) \right\} \langle O_2\rangle
    + \frac{\alpha_s}{4\pi\varepsilon} \left(
    \frac{3N}{8}\,\langle O_1\rangle
    + \frac{2C_F}{3}\,\langle O_4\rangle
    + \frac{C_F}{6}\,\langle O_5\rangle \right) \,,
\end{eqnarray}
where $N$ is the number of colours, and $C_F=\frac 12(N-1/N)$ is the
eigenvalue of the quadratic Casimir operator in the fundamental
representation. Eliminating the operators $O_4$ and $O_5$ by means of
the relations $\langle O_4\rangle=-\frac 12\langle O_1\rangle$ and
$\langle O_5\rangle=-\langle O_2\rangle$, we obtain
\begin{eqnarray}
   \langle O_1^{\rm bare}\rangle &=& \left\{ 1
    + \frac{\alpha_s}{4\pi\varepsilon} \left(
    - \frac{N}{2} - \frac{C_F}{2} \right) \right\} \langle O_1\rangle
    + \frac{\alpha_s}{4\pi\varepsilon}\,N\,\langle O_2\rangle \,,
    \nonumber\\
   \langle O_2^{\rm bare}\rangle &=& \left\{ 1
    + \frac{\alpha_s}{4\pi\varepsilon} \left(
    - \frac{3N}{2} + \frac{5 C_F}{6} \right) \right\}
    \langle O_2\rangle
    + \frac{\alpha_s}{4\pi\varepsilon} \left(
    \frac{3N}{8} - \frac{C_F}{3}\right) \langle O_1\rangle \,.
\end{eqnarray}
These results define a matrix $Z_{ij}$ of renormalization constants
through $\langle O_i^{\rm bare}\rangle = Z_{ij} \langle O_j\rangle$.
Denoting by $Z_{ij}^{(1)}$ the coefficient of the $1/\varepsilon$
pole in this matrix and using the relation \cite{Flor}
\begin{equation}
   \gamma_{ij} = -2\alpha_s\frac{\partial}{\partial\alpha_s}\,
   Z_{ij}^{(1)} \,,
\end{equation}
we obtain the anomalous dimensions appearing in the RGE for the
renormalized operators:
\begin{equation}
   \mu\frac{{\rm d}}{{\rm d}\mu}\,O_i + \gamma_{ij} O_j = 0 \,.
\label{RGE}
\end{equation}
At the one-loop order, the result reads
\begin{equation}
   \hat\gamma = \frac{\alpha_s}{4\pi} \left(
   \begin{array}{ccc}
   N + C_F & ~~ & - 2 N \\
   & & \\
   - \frac 34 N + \frac 23 C_F & ~~ & 3 N - \frac 53 C_F \\
   \end{array}
   \right) \,.
\label{gamma}
\end{equation}

It is remarkable that the eigenvalues of the one-loop anomalous
dimension matrix are given by irrational numbers (in units of
$\alpha_s/4\pi$):
\begin{equation}
   \gamma_\pm = \left( 2 N - \frac{C_F}{3} \right) \pm
   \sqrt{ \frac{5 N^2}{2} - 4 N C_F + \frac{16 C_F^2}{9}}
   = \frac 19 \left( 50 \pm \sqrt{\frac{1565}{2}} \right) \,.
\label{gampm}
\end{equation}
We know of no other case where this happens at the one-loop order. In
the leading logarithmic approximation, the solution of the RGE
(\ref{RGE}) is given by
\begin{equation}
   O_i(m_Q) = U_{ij}(m_Q,\mu) O_j(\mu) \,,
\end{equation}
where
\begin{equation}
   \hat U(m_Q,\mu) = \hat V
   \left( \begin{array}{cc}
   r_+~ & 0 \\
   0~ & r_- \\
   \end{array} \right)
   \hat V^{-1} \,, \qquad
   r_\pm = \left( \frac{\alpha_s(m_Q)}{\alpha_s(\mu)}
   \right)^{\gamma_\pm/2\beta_0} \,.
\label{hatU}
\end{equation}
Here $\beta_0 = \frac{11}{3} N - \frac 23 n_f$ is the first
coefficient of the $\beta$ function, and $\hat V$ is the matrix that
diagonalizes the anomalous dimension matrix:
\begin{equation}
   \hat V^{-1}\,\hat\gamma\,\hat V
   = \frac{\alpha_s}{4\pi}
   \left( \begin{array}{cc}
   \gamma_+~ & 0 \\
   0~ & \gamma_- \\
   \end{array} \right) \,.
\end{equation}
The explicit form of the evolution matrix is
\begin{equation}
   \hat U(m_Q,\mu) = \left( \begin{array}{ccc}
   \frac 12(r_+ + r_-) - (N-\frac 43 C_F) \Delta & ~~ &
    -2 N \Delta \\
   & & \\
   (-\frac 34 N+\frac 23 C_F) \Delta & ~~ & \frac 12(r_+ + r_-)
    + (N-\frac 43 C_F) \Delta \\
   \end{array} \right) \,,
\label{Uexpl}
\end{equation}
where $\Delta=(r_+ - r_-)/(\gamma_+ - \gamma_-)$.

\section{Complete operator basis}
\label{sec:3}

We now consider the general case where the total momentum carried by
the operators does not vanish. The purpose of this section is to show
that even then the operator basis $(O_1,O_2)$ closes under
renormalization. The reader not concerned about this issue can
proceed directly with Sec.~\ref{sec:4}.

To obtain a complete basis of local, scalar (or pseudoscalar)
dimension-5 operators, we have to consider, in addition to
(\ref{ops}), operators with one or two derivatives acting on the
light-quark field. These differ from the operators considered so far
by total derivatives. The number of such operators is reduced
significantly when the equations of motion are used. We find that a
complete basis contains only four new operators in addition to $O_1$
and $O_2$, and choose them in the following form:
\begin{eqnarray}
   T_1 &=& i\partial_\mu (\bar q\,iD^\mu h_v) \,, \nonumber\\
   T_2 &=& (iv\!\cdot\!\partial)^2 (\bar q\,h_v) \,, \nonumber\\
   T_3 &=& (i\partial)^2 (\bar q\,h_v) \,, \nonumber\\
   T_4 &=& i\partial_\mu\,iv\!\cdot\!\partial\,
    (\bar q\,\gamma^\mu h_v) \,.
\end{eqnarray}
In momentum space, the total derivative of a local operator
corresponds to the total external momentum carried by that operator
and thus does not affect the behaviour under renormalization.
Therefore, as far as the calculation of ultraviolet divergences is
concerned, $T_1$ behaves like a dimension-4 operator, while $T_2$,
$T_3$ and $T_4$ behave like dimension-3 operators. Under
renormalization, operators of lower ``effective dimension'' cannot
mix into higher-dimensional operators, but the contrary is, in
general, not true. As a consequence, the Wilson coefficients of the
operators $O_1$ and $O_2$ could, in principle, be modified by the
presence of the lower-dimensional operators.

It follows from this discussion that the anomalous dimension matrix
governing the mixing of the operators $O_i$ and $T_i$ is of the
general form
\begin{equation}
   \hat\Gamma = \left(
   \begin{array}{cc}
   \hat\gamma & ~\hat A \\
   \hat 0 & ~\hat B \\
   \end{array}
   \right) \,,
\end{equation}
where $\hat\gamma$ is the $2\times 2$ matrix given (at the one-loop
order) in (\ref{gamma}). The $4\times 4$ matrix $\hat B$ describing
the mixing of the operators $T_i$ among themselves has been
calculated in Ref.~\cite{MN94}. To all orders in perturbation theory,
it has the simple form
\begin{equation}
   \hat B = \gamma_{\rm hl}\,\hat 1 + \gamma^a \left(
   \begin{array}{cccc}
   1\phantom{-} & \frac 23 & -1\phantom{-} & \frac 13 \\
   0\phantom{-} & 0 & 0 & 0 \\
   0\phantom{-} & 0 & 0 & 0 \\
   0\phantom{-} & 0 & 0 & 0 \\
   \end{array}
   \right) \,,
\end{equation}
where $\gamma_{\rm hl}$ is the anomalous dimension of heavy--light
current operators of dimension 3 \cite{KoRa}--\cite{Gim}, and
$\gamma^a$ has been defined in Ref.~\cite{MN94}. At the one-loop
order, one finds $\gamma_{\rm hl}=-\gamma^a=\gamma_0(\alpha_s/4\pi)$,
where
\begin{equation}
   \gamma_0 = - 3 C_F \,.
\label{gam0}
\end{equation}

\begin{figure}[htb]
\epsfysize=2cm
\centerline{\epsffile{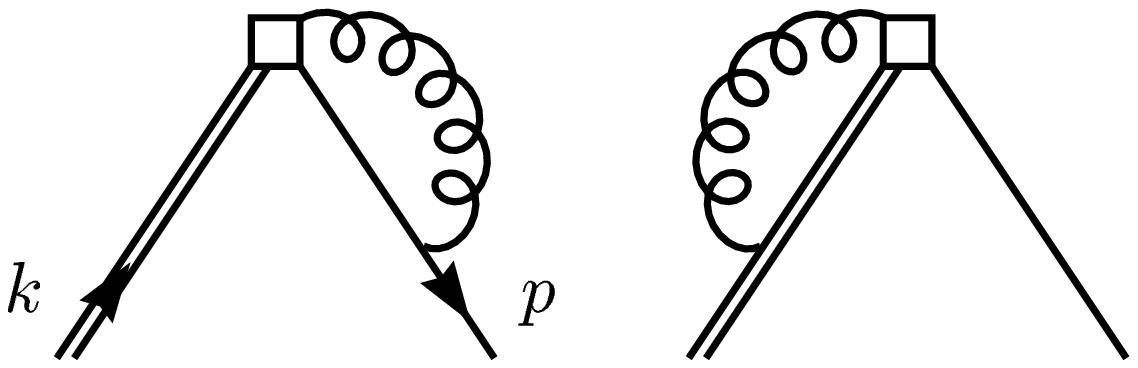}}
\centerline{\parbox{14cm}{\caption{\label{fig:2point}
One-loop diagrams contributing to the calculation of the anomalous
dimension matrix $\hat A$.}}}
\end{figure}

It remains to find the $2\times 4$ matrix $\hat A$ describing the
mixing of $O_1$ and $O_2$ into the operators $T_i$. Since these
operators contain total derivatives, the calculation must be
performed with non-zero total momentum. However, a simplification is
that the operators $T_i$ have non-vanishing quark matrix elements at
tree level. The matrix $\hat A$ can thus be calculated by considering
a two-point (rather than three-point) vertex function. The relevant
one-loop diagrams are shown in Fig.~\ref{fig:2point}. For arbitrary
quark momenta, we find the pole terms:
\begin{eqnarray}
   \langle O_1^{\rm bare}\rangle &=& C_F\,
    \frac{\alpha_s}{4\pi\varepsilon}\,\bar u_q(p)\,(18 p^2)\,
    u_h(v,k) + \dots \,, \nonumber\\
   \langle O_2^{\rm bare}\rangle &=& C_F\,
    \frac{\alpha_s}{4\pi\varepsilon}\,\bar u_q(p)\,
    (4 p^2 + 2 v\cdot p\,\rlap/p)\,u_h(v,k) + \dots \,.
\end{eqnarray}
Both matrix elements vanish on-shell, implying that there is no
mixing of the dimension-5 operators $O_1$ and $O_2$ into the
operators $T_i$, i.e.
\begin{equation}
   \hat A = 0 \,, \qquad
   \hat\Gamma = \left(
   \begin{array}{cc}
   \hat\gamma & ~\hat 0 \\
   \hat 0 & ~\hat B \\
   \end{array}
   \right)\,.
\end{equation}
In other words, the evolution of the operators $T_i$ is disconnected
from that of the operators $O_i$. Moreover, we shall see below that
(at least to next-to-leading order) the operators $T_i$ are not
induced in the matching of QCD operators onto HQET operators. This is
a virtue of our particular choice of the operators in (\ref{ops})
containing derivatives acting only on the heavy-quark field. Hence,
from now on the operators $T_i$ can be omitted from our discussion.

\section{Hybrid renormalization of penguins}
\label{sec:4}

We now include in our discussion local operators containing the
photon field. This is most conveniently done by extending the
definition of the covariant derivative ($D^\mu=\partial^\mu - i g_s
t_a A_a^\mu - i e {\cal A}^\mu$), so that $[iD^\mu,iD^\nu]=ig_s
G^{\mu\nu} + ie F^{\mu\nu}$, where $F^{\mu\nu}$ is the
electromagnetic field. Then the two gluon operators $O_1$ and $O_2$
in (\ref{ops}) are supplemented by their photon counterparts:
\begin{eqnarray}
   O_1^\gamma &=& e\,\bar q\,\sigma_{\mu\nu} F^{\mu\nu} h_v \,,
    \nonumber\\
   O_2^\gamma &=& e\,\bar q\,\gamma_\mu v_\nu\,iF^{\mu\nu} h_v \,.
\end{eqnarray}
The problem is to find the mixing of these four operators under
renormalization.

\begin{figure}[htb]
\epsfxsize=11cm
\centerline{\epsffile{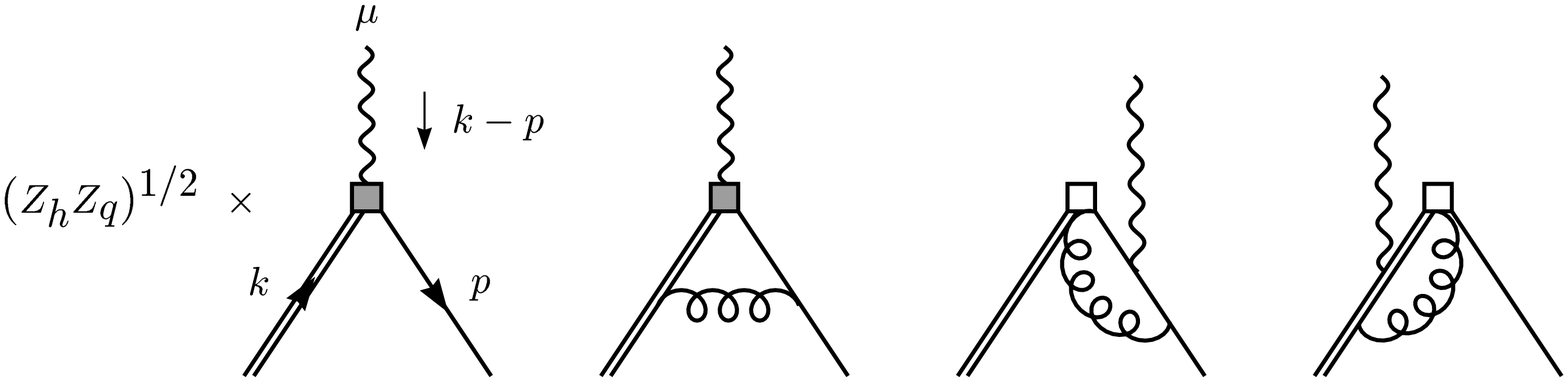}}
\centerline{\parbox{14cm}{\caption{\label{fig:diagrams_gam}
Additional one-loop diagrams contributing to the calculation of the
anomalous dimensions in the presence of photon penguin operators,
denoted by the gray square.}}}
\end{figure}

At leading order in the fine-structure constant $\alpha$, the gluon
operators can mix into the photon operators, but not vice versa.
Thus, we write the corresponding $4\times 4$ anomalous dimension
matrix in the form
\begin{equation}
   \hat\Gamma = \left(
   \begin{array}{cc}
   \hat\gamma~ & \hat X \\
   \hat 0~ & \hat Y \\
   \end{array}
   \right) \,.
\label{Gam}
\end{equation}
The $2\times 2$ submatrices $\hat X$ and $\hat Y$ can be obtained
from the calculation of the diagrams shown in
Fig.~\ref{fig:diagrams_gam}. It is straightforward to see that the
values of these diagrams can be obtained from the results for the
corresponding diagrams in Fig.~\ref{fig:diagrams} by performing a
simple replacement of colour factors, namely by taking the limit
$N\to 0$ keeping $C_F$ fixed. The first two diagrams in
Fig.~\ref{fig:diagrams_gam} determine the matrix $\hat Y$ and yield
\begin{equation}
   \hat Y = \gamma_{\rm hl}\,\hat 1
   = \frac{\alpha_s}{4\pi}\,(-3 C_F)\,\hat 1 \,.
\end{equation}
Since QCD is blind to the photon field, the operators $O_3$ and $O_4$
renormalize as dimension-3 heavy--light current operators, i.e.\
their anomalous dimension is given by $\gamma_{\rm hl}$. The
submatrix $\hat X$ is obtained from the third diagram in
Fig.~\ref{fig:diagrams_gam}, whereas the last diagram vanishes. The
result is
\begin{equation}
   \hat X = \frac{\alpha_s}{4\pi}\,Q_q C_F \left(
   \begin{array}{ccc}
   4 & ~ & 0 \\
   \frac 23 & ~ & \frac 43 \\
   \end{array}
   \right) \,,
\label{hatX}
\end{equation}
where $Q_q$ is the electric charge of the light quark in units of
$e$. Note that the sum of these two matrices reproduces the ``abelian
part'' of the matrix $\hat\gamma$ in (\ref{gamma}), i.e.\ $\hat X/Q_q
+ \hat Y = \lim_{N\to 0}\,\hat\gamma$.

The three eigenvalues of the anomalous dimension matrix $\hat\Gamma$
in units of $\alpha_s/4\pi$ are $\gamma_\pm$ in (\ref{gampm}) and
$\gamma_0$ in (\ref{gam0}). The solution of the RGE for the four
operators $O_i$ can be obtained either by diagonalizing the $4\times
4$ matrix $\hat\Gamma$ directly, or by solving first the homogeneous
equation for the two operators $O_1^\gamma$ and $O_2^\gamma$, and
then inserting the solution for these operators into the
inhomogeneous equation for the operators $O_1$ and $O_2$. In leading
logarithmic order, we find that the photon operators renormalize
multiplicatively:
\begin{equation}
   O_i^\gamma(m_Q) = r_0\,O_i^\gamma(\mu) \,;\quad i=1,2,
\label{RGEsol1}
\end{equation}
with
\begin{equation}
   r_0 = \left( \frac{\alpha_s(m_Q)}{\alpha_s(\mu)}
   \right)^{\gamma_0/2\beta_0} \,.
\label{r0}
\end{equation}
The evolution of the gluon operators is given by
\begin{equation}
   \left( \begin{array}{c} O_1 \\ O_2 \\ \end{array} \right)_{\!m_Q}
   = \hat U(m_Q,\mu)
   \left( \begin{array}{c} O_1 \\ O_2 \\ \end{array} \right)_{\!\mu}
   + \hat W(m_Q,\mu)\,\hat X_{(0)}
   \left( \begin{array}{c} O_1^\gamma \\ O_2^\gamma \\ \end{array}
   \right)_{\!\mu} \,,
\label{RGEsol2}
\end{equation}
where $\hat X_{(0)}$ denotes the matrix $\hat X$ in units of
$\alpha_s/4\pi$, the evolution matrix $\hat U$ has been given in
(\ref{hatU}) and (\ref{Uexpl}), and
\begin{equation}
   \hat W(m_Q,\mu) = \hat V
   \left( \begin{array}{ccc}
   \displaystyle\frac{r_+ - r_0}{\gamma_+ - \gamma_0} & ~ & 0 \\
   0 & ~ & \displaystyle\frac{r_- - r_0}{\gamma_- - \gamma_0} \\
   \end{array} \right)
   \hat V^{-1} \,.
\end{equation}

Consider now the effective Hamiltonian in (\ref{Heff}). At the scale
$m_Q$, it contains the gluon and photon penguin operators
\begin{equation}
   {\cal Q}_g = g_s\,\bar q\,\sigma_{\mu\nu} G^{\mu\nu} Q \,, \qquad
   {\cal Q}_\gamma = e\,\bar q\,\sigma_{\mu\nu} F^{\mu\nu} Q \,,
\label{Qg}
\end{equation}
with known coefficients $c_g(m_Q)$ and $c_\gamma(m_Q)$. Once again
$q$ may be replaced by $q_L$ or $q_R$. Our goal is to evolve the
Hamiltonian down to a low renormalization scale $\mu\ll m_Q$. Each of
the two penguin operators (renormalized at the scale $m_Q$) has an
expansion in terms of the HQET operators renormalized at a scale
$\mu<m_Q$. We define a set of coefficient functions by:
\begin{eqnarray}
   {\cal Q}_g(m_Q) &\to& O_1(m_Q) = \sum_{i=1,2}\,\left[
    C_i^g(\mu)\,O_i(\mu) + C_i^\gamma(\mu)\,O_i^\gamma \right] \,,
    \nonumber\\
   {\cal Q}_\gamma(m_Q) &\to& O_1^\gamma(m_Q) = \sum_{i=1,2}\,
    D_i^\gamma(\mu)\,O_i^\gamma(\mu) \,.
\label{CD}
\end{eqnarray}
At leading logarithmic order, the initial values of the coefficients
at the scale $m_Q$ are determined by a tree-level comparison of
operators matrix elements in QCD and in the HQET and are thus simply
given by $C_1^g(m_Q)=D_1^\gamma(m_Q)=1$; all other coefficients
vanish at the scale $m_Q$. The values of the coefficients at a lower
scale $\mu<m_Q$ can be read of from the solution of the RGE in
(\ref{RGEsol1}) and (\ref{RGEsol2}). We obtain
\begin{equation}
   D_1^\gamma(\mu) = r_0 \,, \qquad D_2^\gamma(\mu) = 0 \,,
\end{equation}
and
\begin{eqnarray}
   C_1^g(\mu) &=& \frac{r_+ + r_-}{2} + \frac 12
    \left( 1 - \frac{4 C_F}{3 N} \right) C_2^g(\mu) \,, \nonumber\\
   C_2^g(\mu) &=& - 2 N\,\frac{r_+ - r_-}{\gamma_+ - \gamma_-} \,,
    \nonumber\\
   C_1^\gamma(\mu) &=& 2 Q_q C_F \left\{
    \frac{r_+ - r_0}{\gamma_+ - \gamma_0}
    + \frac{r_- - r_0}{\gamma_- - \gamma_0} \right\}
    + 2 \left( 1 - \frac{C_F}{N} \right) C_2^\gamma(\mu) \,,
    \nonumber\\
   C_2^\gamma(\mu) &=& - \frac 83\,Q_q\,
    \frac{N C_F}{\gamma_+ - \gamma_-} \left\{
    \frac{r_+ - r_0}{\gamma_+ - \gamma_0}
    - \frac{r_- - r_0}{\gamma_- - \gamma_0} \right\} \,.
\end{eqnarray}
With $N=3$ and $C_F=4/3$, this yields:
\begin{eqnarray}
   C_1^g(\mu) &\simeq& 0.6966\,r_- + 0.3034\,r_+ \,, \nonumber\\
   C_2^g(\mu) &\simeq& 0.9652\,(r_- - r_+) \,, \nonumber\\
   C_1^\gamma(\mu) &\simeq& Q_q\,(0.7093\,r_- + 0.0600\,r_+
    - 0.7693\,r_0) \,, \nonumber\\
   C_2^\gamma(\mu) &\simeq& Q_q\,(0.2661\,r_- - 0.1355\,r_+
    - 0.1306\,r_0) \,.
\label{numbers}
\end{eqnarray}
As an example, we evaluate the coefficients for the $b$ quark at the
scale $\mu=1$~GeV, using $n_f=4$ flavours (i.e.\ $\beta_0=25/3$),
$\alpha_s(m_b)=0.210$ and $\alpha_s(\mu)=0.458$. This gives
$D_1^\gamma\simeq 1.21$, as well as $C_1^g\simeq 0.82$, $C_2^g\simeq
0.22$, $C_1^\gamma\simeq -0.26\,Q_q$, and $C_2^\gamma\simeq
-0.01\,Q_q$. We observe that the effects of hybrid renormalization
are typically of order 20\%, except for the coefficient $C_2^\gamma$,
which is of order $[\alpha_s\ln(m_Q/\mu)]^2$.

\section{One-loop matching}
\label{sec:5}

In the above discussion of operator evolution, we have combined the
tree-level matching of the QCD penguin operators onto the HQET
penguin operators at the scale $m_b$ with the evolution equations
solved in the leading logarithmic approximation. This approach is
justified if $\ln(m_Q/\mu)\gg 1$. As an alternative, we shall now
discuss the full one-loop matching of the QCD operators onto the HQET
operators. Our calculation will be equivalent to that performed by
Eichten and Hill for the chromo-magnetic operator containing two
heavy-quark fields \cite{EiHi}. The advantage of this approach is
that we will be able to include non-logarithmic terms of
$O(\alpha_s)$. On the other hand, we will not be able to resum
logarithmic terms to higher orders in perturbation theory. A
consistent combination of one-loop matching with
renormalization-group improvement would require to calculate the
operator anomalous dimensions to two-loop order, which is beyond the
scope of our work. However, once this calculation will have been
done, the one-loop matching conditions at the scale $m_Q$ derived
here will be part of the full next-to-leading order analysis.

To derive expressions for the Wilson coefficients, we must compare
the
matrix elements of the QCD operators ${\cal Q}_g$ and ${\cal
Q}_\gamma$ in (\ref{Qg}) with the corresponding matrix elements of
the HQET operators $O_i$ and $O_i^\gamma$. Order by order in
perturbation theory, the comparison (``matching'') of these matrix
elements defines the coefficient functions. By construction, the
differences between matrix elements in the two theories are
insensitive to any long-distance properties, such as the nature of
the infrared regulator or of the external states. Therefore, it is
legitimate to perform the matching calculation using (on-shell) quark
and gluon states, and working with any infrared regularization scheme
that is convenient. Following Refs.~\cite{EiHi,MN94}, we choose to
regulate both ultraviolet and infrared divergences using dimensional
regularization. Moreover, we expand the resulting expressions for the
Feynman amplitudes in powers of the external momenta, and then set
the external momenta to zero inside the loop integrals. Then the only
mass scale remaining in the QCD calculation is the heavy-quark mass,
and hence only diagrams containing a heavy-quark propagator in a loop
contribute. Likewise, in the HQET there is no mass scale left after
the external momenta are set to zero, and hence all loop integrals
vanish on dimensional grounds. So only the tree-level matrix elements
of the HQET operators multiplied by their Wilson coefficient
functions remain. That is why this particular regularization scheme
is most economic for our purpose.

In the matching calculation, we have to consider all operators in the
HQET that have the same quantum numbers as the QCD operator in
(\ref{Qg}). In addition to the genuine dimension-5 operators $O_i$
and $O_i^\gamma$ and the total derivatives $T_i$ encountered so far,
this set includes in principle also operators proportional to the
heavy-quark mass $m_Q$. They are:
\begin{eqnarray}
   S_1 &=& m_Q\,iv\!\cdot\!\partial\,(\bar q\,h_v) \,, \nonumber\\
   S_2 &=& m_Q\,i\partial_\mu (\bar q\,\gamma^\mu h_v) \,,
\nonumber\\
   S_3 &=& m_Q\,\bar q\,iv\!\cdot\!D\,h_v \,, \nonumber\\
   S_4 &=& m_Q\,\bar q\,i\rlap{\,/}D h_v \,, \nonumber\\
   S_5 &=& m_Q^2\,\bar q\,h_v \,.
\end{eqnarray}
The matrix elements of the operators $S_3$ and $S_4-S_2$ evaluated
between physical states vanish by the equations of motion.
Nevertheless, these operators may be induced when the matching
calculation is performed with unphysical states, such as on-shell
quarks and gluons.

\begin{figure}[htb]
\epsfxsize=12cm
\centerline{\epsffile{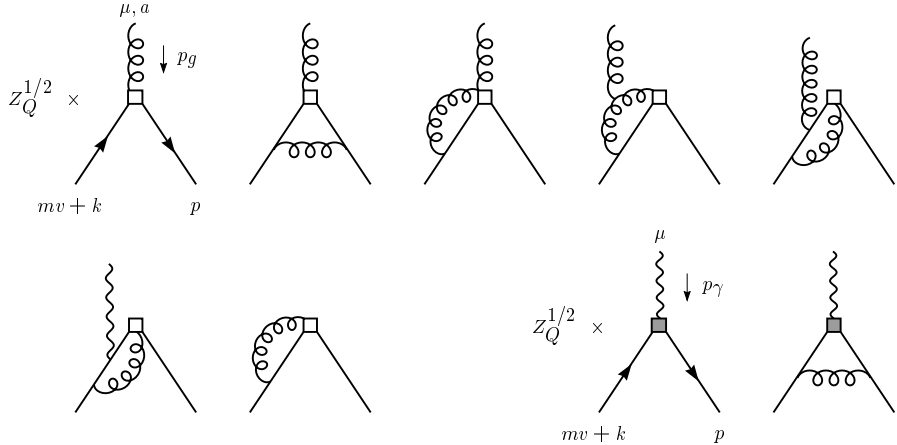}}
\centerline{\parbox{14cm}{\caption{\label{fig:match}
One-loop diagrams contributing to the matching calculation for the
gluon and photon penguin operators, ${\cal Q}_g$ and ${\cal
Q}_\gamma$, denoted by the white and gray squares.}}}
\end{figure}

Since some operators ($T_2$, $T_3$ and $T_4$, as well as $S_1$, $S_2$
and $S_5$) do not have gluon matrix elements at tree level, we have
to perform the matching calculation by considering both the
quark--quark and the quark--quark--gluon vertex functions, with
arbitrary (but on-shell) external momenta. The relevant diagrams are
shown in Fig.~\ref{fig:match}. We have performed the calculation of
these diagrams in an arbitrary covariant gauge, and find that the
results for the coefficient functions are gauge independent. The
on-shell wave-function renormalization constant for the heavy-quark
field in our regularization scheme is
\begin{equation}
   Z_Q = 1 - C_F\,\frac{\alpha_s}{4\pi} \left(
   \frac{m_Q^2}{4\pi\mu^2} \right)^{d/2-2}
   \frac{(d-1)\,\Gamma(2-d/2)}{(d-3)} \,,
\end{equation}
which is also gauge independent. The light-quark field is not
renormalized in this scheme (i.e.\ $Z_q=1$). Note that only the
second diagram in Fig.~\ref{fig:match} involves a light-quark
propagator in a loop. It is easily seen that the contribution of this
diagram is linear in the external gluon momentum, but independent of
the momentum of the light quark. Hence, it follows that no
contributions proportional to the light-quark momentum appear in the
matching calculation. This means that the operators $T_i$ and $S_1$,
$S_2$ are not induced by radiative corrections, at least not to
$O(\alpha_s)$.

Let us now present the result of the evaluation of the diagrams
involving the gluon penguin operator ${\cal Q}_g$. For the sum of all
contributions, we find
\begin{eqnarray}
   \langle {\cal Q}_g\rangle &=& C_1^g\,\langle O_1\rangle
    + C_2^g\,\langle O_2\rangle + C_1^\gamma\,\langle O_1^\gamma
    \rangle + C_2^\gamma\,\langle O_2^\gamma\rangle \nonumber\\
   &&\mbox{}+ C_3^g\,\langle O_3\rangle
    + C_S\,\left( - 2 \langle S_3\rangle
    + \langle S_4\rangle + 2\langle S_5\rangle \right) \,,
\label{match}
\end{eqnarray}
where the exact expressions for the coefficients in an arbitrary
space-time dimension $d$ are
\begin{eqnarray}
   C_1^g &=& 1 + \frac{K}{2} \left[
    (2 d^3 - 23 d^2 + 71 d - 62) C_F
    - (d^3 - 10 d^2 + 25 d - 14) N \right] \,, \nonumber\\
   \phantom{ \bigg[ }
   C_2^g &=& - K \left[ 2(d-2)(d-3)(d-4) C_F
    - (d^3 - 9 d^2 + 23 d - 14) N \right] \,, \nonumber\\
   \phantom{ \bigg[ }
   C_1^\gamma &=& K Q_q C_F\,(d-3)(d^2-9 d+16) \,, \nonumber\\
   C_2^\gamma &=& -2 K Q_q C_F\,(d-4)(d^2-6 d+10) \,,
\end{eqnarray}
as well as
\begin{eqnarray}
   C_3^g &=& -4 K C_F\,(d-1)(d-4) \,, \nonumber\\
   C_S &=& -2 K C_F\,(d-1)(d-3) \,,
\end{eqnarray}
where we have abbreviated
\begin{equation}
   K = \frac{\alpha_s}{4\pi} \left( \frac{m_Q^2}{4\pi\mu^2}
   \right)^{d/2-2} \frac{\Gamma(2-d/2)}{(d-2)(d-3)} \,.
\end{equation}

Note that the operators $O_3$ and $S_3$ induced by one-loop matching
in (\ref{match}) vanish by the equations of motion. These operators
appear only because the matching calculation is (legitimately) being
performed using unphysical quark and gluon states, rather than
physical hadron states. The quark states satisfy the on-shell
conditions $v\cdot k=0$ and $\rlap/p=p^2=0$, which differ from the
true equations of motion at $O(g_s)$. The presence of operators that
vanish by the equations of motion in (\ref{match}) is, however,
without physical significance. For all practical applications these
operators can be ignored, since their physical matrix elements
vanish. A similar statement obtains for the operators $S_4$ and
$S_5$. That fact that they appear in (\ref{match}) in the combination
$S_5+\frac12 S_4$ is a consequence of the so-called reparametrization
invariance of the HQET \cite{LuMa}. Indeed, this combination is
nothing but the HQET counterpart of the QCD operator $m_Q^2\,\bar
q\,Q$. But this operator has the form of an off-diagonal mass term,
which has no observable effect as it can be removed by a redefinition
of the quark fields. This simply shifts the physical quark masses by
an amount of $O(G_F\alpha_s m_Q^3)$. This discussion shows {\it a
posteriori\/} that, as shown in (\ref{CD}), the matching of the QCD
operator ${\cal Q}_g$ onto the HQET operators involves only genuine
dimension-5 operators, preserving thus the structure of an effective
field theory, in which operators enhanced by powers of the large
scale $m_Q$ do not contribute to physical matrix elements.

A similar calculation for the matrix element of the photon penguin
operator gives, omitting now unphysical operators,
\begin{equation}
   \langle {\cal Q}_\gamma\rangle = D_1^\gamma\,\langle O_1^\gamma
   \rangle + D_2^\gamma\,\langle O_2^\gamma\rangle \,,
\end{equation}
where
\begin{eqnarray}
   D_1^\gamma &=& 1 + \frac{K}{2}\,C_F\,(d^2-15 d+34) \,, \nonumber\\
   D_2^\gamma &=& -2 K C_F\,(d-4)^2 \,.
\end{eqnarray}

Let us now set $d=4-2\varepsilon$, take the limit $\varepsilon\to 0$
and remove the poles in $1/\varepsilon$ using the
$\overline{\mbox{\sc ms}}$ subtraction prescription. In doing this,
we have to take account of the fact that not only the matrix elements
of the HQET operators are ultraviolet divergent, but also the matrix
elements of the QCD operators ${\cal Q}_g$ and ${\cal Q}_\gamma$
themselves. If we are interested in the evolution of the operators
below the scale $m_Q$, taking their values at $\mu=m_Q$ as given, we
have to remove this second type of divergence by renormalizing the
bare QCD operators at the scale $\mu=m_Q$. This is accomplished by a
$2\times 2$ matrix $\hat Z^{-1}$, which is given by
\cite{GSW,GDSN}:\footnote{The matrix in parenthesis contains the
one-loop anomalous dimensions of the QCD penguin operators, after a
factor of the heavy-quark mass has been removed.}
\begin{equation}
   \hat Z^{-1} = \hat 1 + \frac{\alpha_s}{8\pi} \left(
   \frac{m_Q^2}{4\pi\mu^2} \right)^{d/2-2} \Gamma(2-d/2)
   \left( \begin{array}{ccc}
   2 C_F & ~~ & 0 \\
   & & \\
   8 Q_q C_F  & ~~ & 10 C_F - 4 N \\
   \end{array} \right) \,,
\end{equation}
which multiplies the $2\times 4$ matrix of coefficient functions
\begin{equation}
   \hat C = \left( \begin{array}{cccc}
   C_1^g & C_2^g & C_1^\gamma & C_2^\gamma \\
   & & & \\
   0 & 0 & D_1^\gamma & D_2^\gamma \\
   \end{array} \right)
\end{equation}
from the left. The remaining divergences are removed by subtracting
the $1/\varepsilon$ poles using the $\overline{\mbox{\sc ms}}$
scheme. The result is:
\begin{equation}
   D_1^\gamma(\mu) = 1 - C_F\,\frac{\alpha_s}{\pi} \left(
    \frac 38\ln\frac{\mu^2}{m_Q^2} + 1 \right) \,, \qquad
   D_2^\gamma(\mu) = 0 \,,
\end{equation}
and
\begin{eqnarray}
   C_1^g(\mu) &=& 1 + (N + C_F)\,\frac{\alpha_s}{8\pi}\,
    \ln\frac{\mu^2}{m_Q^2} + \left( N - \frac 54 C_F \right)
    \frac{\alpha_s}{\pi} \,, \nonumber\\
   C_2^g(\mu) &=& - N\,\frac{\alpha_s}{4\pi}\,\ln\frac{\mu^2}{m_Q^2}
    - (N - 2 C_F)\,\frac{\alpha_s}{2\pi} \,, \nonumber\\
   C_1^\gamma(\mu) &=& Q_q C_F\,\frac{\alpha_s}{2\pi} \left(
    \ln\frac{\mu^2}{m_Q^2} - \frac 12 \right) \,, \nonumber\\
   C_2^\gamma(\mu) &=& Q_q C_F\,\frac{\alpha_s}{\pi} \,.
\label{coefs}
\end{eqnarray}
The Wilson coefficients obey the RGE
\begin{equation}
   \mu\frac{{\rm d}}{{\rm d}\mu}\,\hat C(\mu)
   = \hat C(\mu)\,\hat\Gamma \,,
\end{equation}
where $\hat\Gamma$ has been given in (\ref{Gam})--(\ref{hatX}). The
fact that our explicit results for the coefficient functions satisfy
this equation is a strong check of our results.

As at the end of Sec.~\ref{sec:4}, we now evaluate as an example the
coefficients for the $b$ quark at the scale $\mu=1$~GeV, using
$m_b=4.8$~GeV and $\alpha_s=\alpha_s(\sqrt{\mu\,m_b})=0.282$. This
gives $D_1^\gamma\simeq 1.02$, as well as $C_1^g\simeq 0.97$,
$C_2^g\simeq 0.20$, $C_1^\gamma\simeq -0.22\,Q_q$, and
$C_2^\gamma\simeq 0.12\,Q_q$. The comparison with the leading-order
renormalization-group improved results presented after
(\ref{numbers}) gives an idea of the importance of next-to-leading
corrections. In some cases, such as $D_1^\gamma$ and $C_2^\gamma$,
the non-logarithmic terms of $O(\alpha_s)$ in (\ref{coefs}) are quite
important. For those coefficients a full next-to-leading order
calculation would be desirable.

\section{Meson decay constants and wave functions}
\label{sec:6}

Another application of our results is the calculation of higher-order
corrections in the heavy-quark expansion of current matrix elements.
As an example, we discuss the matrix elements of heavy--light
currents between a ground-state meson and the vacuum, which define
the meson decay constants $f_M$. The heavy-quark expansion for
heavy-meson decay constants has been discussed at next-to-leading
order in $1/m_Q$ in Ref.~\cite{subl}. In general, at order $1/m_Q^n$
in the expansion there appear contributions from the matrix elements
of local current operators of dimension $3+n$, and (for $n\ge 1$)
from the matrix elements of non-local operators containing the
time-ordered products of lower-dimensional current operators with
terms from the effective Lagrangian of the HQET. Our results are
relevant to the discussion of the local corrections of order
$1/m_Q^2$. They can be expressed in terms of the matrix elements of
operators of the type\footnote{There are also contributions from
operators that are total derivatives; however, their matrix elements
can be expressed in terms of parameters already encountered at order
$1/m_Q$ in the expansion.}
$\bar q\,\Gamma\,iD^\mu iD^\nu h_v$. Unlike the case considered so
far in this paper, these operators transform as vectors and axial
vectors under the Lorentz group. However, our results still apply
because of heavy-quark spin symmetry. In fact, the most general
matrix element of operators of the type shown above can be written in
the form
\begin{equation}
   \langle\,0\,|\,\bar q\,\Gamma\,iD^\mu iD^\nu h_v\,|M(v)\rangle_\mu
   = \frac 12\sqrt{m_M}\,F(\mu)\,\mbox{Tr}\left\{
   \Theta^{\mu\nu}\,\Gamma\,{\cal M}(v) \right\} \,,
\label{trace}
\end{equation}
where $F\approx f_M\sqrt{m_M}$ is the leading contribution to the
meson decay constant, and
\begin{equation}
   {\cal M}(v) = \frac{1+\rlap/v}{2}\left\{ \begin{array}{ll}
   -i\gamma_5 & \mbox{; pseudoscalar meson $M(v)$,} \\
   \rlap/e & \mbox{; vector meson $M^*(e,v)$,} \\
   \end{array} \right.
\end{equation}
is a Dirac matrix representing the spin wave function of the
ground-state mesons in the HQET \cite{Falk}. Here $v$ is the meson
velocity, and $e$ is the polarization vector of the vector meson
($e\cdot v=0$). The matrix ${\cal M}(v)$ satisfies ${\cal
M}(v)\,\rlap/v = -{\cal M}(v)$. The most general decomposition of the
tensor form factor $\Theta^{\mu\nu}$ consistent with Lorentz
covariance, heavy-quark symmetry and the equations of motion involves
the ``binding energy'' $\bar\Lambda=m_M-m_Q$, which is a
scale-independent mass parameter of the HQET, as well as two new
parameters $\lambda_E^2$ and $\lambda_H^2$ \cite{GrNe}:
\begin{eqnarray}
   \Theta^{\mu\nu} &=& 2 \left[ \bar\Lambda^2 + \lambda_H^2(\mu)
    + \lambda_E^2(\mu) \right]\,(v^\mu v^\nu - g^{\mu\nu})
    \nonumber\\
   &&\mbox{}+ 2\lambda_E^2(\mu)\,v^\mu(\gamma^\nu+v^\nu)
    + \frac i2\,\lambda_H^2(\mu)\,\sigma^{\mu\nu}\,(1-\rlap/v) \,.
\end{eqnarray}
Heavy-quark spin symmetry ensures that the relation (\ref{trace})
holds for any matrix $\Gamma$. Thus, we are free to set
$\Gamma=\gamma_5(\gamma_\mu v_\nu - \gamma_\nu v_\mu)$ or
$\Gamma=i\gamma_5\sigma_{\mu\nu}$, in which case we find relations
between the parameters $\lambda_E^2$ and $\lambda_H^2$ and the matrix
elements of the pseudoscalar operators $O_1^{(5)}$ and $O_2^{(5)}$
evaluated between a pseudoscalar meson and the vacuum:
\begin{eqnarray}
   \langle\,0\,|\,O_2^{(5)}\,|M(v)\rangle
   &=& -i\sqrt{m_M}\,F(\mu)\lambda_E^2(\mu) \,, \nonumber\\
   \langle\,0\,|\,\case 12 O_1^{(5)} - O_2^{(5)}\,|M(v)\rangle
   &=& -i\sqrt{m_M}\,F(\mu)\lambda_H^2(\mu) \,.
\end{eqnarray}
In the rest frame of the hadron, these are matrix elements of purely
chromo-electric and chromo-magnetic operators, respectively.

The scale dependence of these matrix elements is determined by the
scale dependence of the operators $O_1$ and $O_2$, which is described
by the evolution matrix given in (\ref{Uexpl}). We obtain
\begin{eqnarray}
   r_0\,\lambda_E^2(m_Q) &=& \left( \frac{r_+ + r_-}{2}
    - \frac{N}{2}\,\frac{r_+ - r_-}{\gamma_+ - \gamma_-} \right)
    \lambda_E^2(\mu) - \left( \frac 32\,N - \frac 43\,C_F \right)
    \frac{r_+ - r_-}{\gamma_+ - \gamma_-}\,\lambda_H^2(\mu)
    \nonumber\\
   &\simeq& (0.7413\,r_- + 0.2587\,r_+)\,\lambda_E^2(\mu)
    + 0.4379\,(r_- - r_+)\,\lambda_H^2(\mu) \,, \nonumber\\
   && \nonumber\\
   r_0\,\lambda_H^2(m_Q) &=& \left( \frac{r_+ + r_-}{2}
    + \frac{N}{2}\,\frac{r_+ - r_-}{\gamma_+ - \gamma_-} \right)
    \lambda_H^2(\mu) - \left( \frac 32\,N - \frac 43\,C_F \right)
    \frac{r_+ - r_-}{\gamma_+ - \gamma_-}\,\lambda_E^2(\mu)
    \nonumber\\
   &\simeq& (0.7413\,r_+ + 0.2587\,r_-)\,\lambda_H^2(\mu)
    + 0.4379\,(r_- - r_+)\,\lambda_E^2(\mu) \,,
\label{lamevol}
\end{eqnarray}
where the factor $r_0$ on the left-hand side compensates for the
scale dependence of the parameter $F(\mu)$, according to the equation
$F(m_Q)=r_0 F(\mu)$, with $r_0$ given in (\ref{r0}).

Let us now discuss the specific example of meson decay constants to
leading logarithmic order. The local corrections in the heavy-quark
expansion are obtained from the expansion of the flavour-changing
current $\bar q\,\gamma^\alpha (1-\gamma_5) Q$, using \cite{review}
\begin{equation}
   Q \to h_v + \frac{i\rlap{\,/}D - iv\!\cdot\!D}{2 m_Q}\,h_v
   + \frac{g_s}{4 m_Q^2}\,\gamma_\mu v_\nu\,iG^{\mu\nu} h_v
   + \dots \,.
\end{equation}
The corresponding expressions for the meson decay constants are
\begin{equation}
   f_M\sqrt{m_M} = F(m_Q) \left\{ 1 - d_M \left(
   \frac{\bar\Lambda}{6 m_Q} + \frac{\lambda_E^2(m_Q)}{12 m_Q^2}
   + \dots \right) + \hbox{non-local terms} \right\} \,,
\label{fM}
\end{equation}
where the spin-dependent coefficient $d_M$ takes the values 3 and
$-1$ for pseudoscalar and vector mesons, respectively. The non-local
matrix elements of order $1/m_Q$ are discussed in Ref.~\cite{subl}.
The new ingredient in (\ref{fM}) is the local correction of order
$1/m_Q^2$, which is determined by the single parameter $\lambda_E^2$
renormalized at the scale $m_Q$. Using our result (\ref{lamevol}),
this parameter can be related to the values of $\lambda_E^2$ and
$\lambda_H^2$ renormalized at some lower scale $\mu$, which is
typically the scale intrinsic to some non-perturbative calculation of
these hadronic parameters. For instance, QCD sum rules have been used
to predict that $\lambda_E^2(\mu)\approx 0.11$~GeV$^2$ and
$\lambda_H^2(\mu)\approx 0.18$~GeV$^2$ at the scale $\mu\approx
1$~GeV \cite{GrNe}. For the $b$ quark, e.g., using the numbers given
at the end of Sec.~\ref{sec:4}, this translates to
$\lambda_E^2(m_b)\approx 0.69\,\lambda_E^2(\mu) +
0.08\,\lambda_H^2(\mu) \approx 0.09$~GeV$^2$, which is the value to
be used in (\ref{fM}). The corresponding correction to the $B$-meson
decay constants is clearly very small, indicating a good convergence
of the local terms in the heavy-quark expansion. Before a meaningful
prediction for the decay constants at order $1/m_Q^2$ can be
obtained, it would however be necessary to include the non-local
terms as well.

For completeness, we note that the hadronic parameters $\lambda_E^2$
and $\lambda_H^2$ are of a more general interest, since they are
related to the second moments of meson wave functions
$\varphi_\pm(\omega,\mu)$. Introducing a vector $z$ on the light-cone
($z^2=0$, $v\cdot z\equiv t$) and working in light-cone gauge
($A_+=0$), we define
\begin{equation}
   \frac{1}{2\pi} \int\mbox{d}t\,e^{i\omega t}\,\langle\,0\,|\,
   \bar q(0)\,\gamma_\pm\Gamma\,h_v(z)\,|M(v)\rangle_\mu
   = \frac 12\sqrt{m_M}\,F(\mu)\,\varphi_\pm(\omega,\mu)\,
   \mbox{Tr}\left\{ \gamma_\pm\Gamma\,{\cal M}(v) \right\} \,,
\end{equation}
where $\gamma_\pm$ are the light-cone projections of the Dirac
matrices. Defining the moments
\begin{equation}
   \langle\omega^n\rangle_\pm = \int\limits_0^\infty\!
   \mbox{d}\omega\,\varphi_\pm(\omega,\mu)\,\omega^n \,,
\end{equation}
which are normalized such that $\langle\omega^0\rangle_\pm=1$, one
can show that \cite{GrNe}\footnote{These results are valid in a
regularization scheme without a dimensionful regulator, such as
dimensional regularization.}
\begin{eqnarray}
   \langle\omega\rangle_\pm &=& \left( 1 \pm \frac 13 \right)
    \bar\Lambda \,, \nonumber\\
   \langle\omega^2\rangle_\pm
   &=& \left( \frac 43 \pm \frac 23 \right) \bar\Lambda^2
    + \left( \frac 13 \pm \frac 13 \right) \lambda_E^2(\mu)
    + \frac 13\,\lambda_H^2(\mu) \,.
\end{eqnarray}
Our results in (\ref{lamevol}) can be used to control the scale
dependence of the second moments.

\section{Conclusions}
\label{sec:7}

We have discussed the hybrid renormalization of local, scalar and
pseudoscalar dimension-5 operators containing a heavy and a light
quark field. Our results determine the scale dependence and mixing of
such operators at scales below the heavy-quark mass. We have
calculated the corresponding anomalous dimensions at the one-loop
order, which is sufficient to obtain the renormalization-group
evolution of the operators in the leading logarithmic approximation.

Two applications of our results have been discussed in detail. The
first one concerns the renormalization of genuine dimension-5 QCD
operators at low renormalization scales. Important examples are the
gluon and photon penguin operators, which appear in the effective
weak Hamiltonian renormalized at the scale $m_b$, and whose evolution
at high scales $\mu\gg m_b$ is well known. We have discussed the
mixing of these penguin operators at a low scale $\mu\ll m_b$ and
derived the effective Hamiltonian for this case. This is relevant for
the calculation of hadronic matrix elements of the penguin operators
performed, e.g., using lattice simulations. In addition to the
evolution at leading logarithmic order, we have also performed the
full one-loop matching of gluon and photon penguin operators onto
their effective-theory counterparts. Besides providing a test of our
results for the operator anomalous dimensions, this calculation will
eventually be part of a full next-to-leading order
renormalization-group improved analysis, once the two-loop anomalous
dimensions of the operators will have been calculated.

The second application concerns the calculation of certain
higher-order corrections in the heavy-quark expansion of current
matrix elements, such as they appear in the description of weak decay
form factors. In particular, local dimension-5 operators appear at
order $1/m_Q^2$ in the heavy-quark expansion of meson decay
constants. We have discussed, as an example, the local $1/m_Q^2$
corrections to the decay constants of heavy mesons at the scale
$m_Q$, at which a single new hadronic parameter appears. Our results
can then be used to evolve the result down to a lower scale $\mu$, at
which non-perturbative evaluations of the relevant hadronic matrix
elements may be performed. A similar analysis could be performed for
semileptonic transition form factors describing, e.g., semileptonic
decays such as $B\to\pi\,\ell\,\nu$.

\vspace{0.3cm}
{\it Acknowledgements:\/}
One of us (M.N.) would like to thank Andrzej Buras, Marco Ciuchini,
Gian Giudice, Howard Haber and Chris Sachrajda for helpful
discussions.

\end{document}